\documentclass{JHEP3}

\input{epsf}
\usepackage{epsfig}

\preprint{ MIT-CTP-3277}

 \title{Winding strings and $AdS_3$ black holes}
 \author{
Jan Troost \\  
    Center for Theoretical Physics \\  MIT \\
    77 Mass Ave \\ Cambridge, MA 02139 \\ USA \\
\email{troost@mit.edu}
   }

 \abstract{We start a systematic study of string theory in $AdS_3$ black
hole backgrounds. Firstly, we analyse in detail the geodesic structure of the
BTZ black hole, including spacelike geodesics. Secondly, we study
the spectrum for massive and massless scalar fields, paying particular 
attention to the connection between $Sl(2,R)$ subgroups, the theory
of special functions and global properties of the BTZ black holes. 
We construct classical strings that wind the black holes.
Finally, we apply the general formalism to 
the vacuum black hole background, and formulate the boundary
spacetime Virasoro algebra in terms of worldsheet operators. We moreover
establish the link between a proposal for a ghost free spectrum for 
$Sl(2,R)$ string
propagation and the massless 
black hole background, thereby claryfing aspects of the
$AdS_3/CFT$ correspondence.}
\begin{document}

\section{Introduction}
In the past ten years, we have learned a lot about string theory
in black hole backgrounds. In \cite{Witten:1991yr} a two-dimensional
black hole background was analysed, using an exact coset conformal
field theory. It was shown that the causal structure of the spacetime
mimicked well its four-dimensional analogue. Using the powerful tools
of two-dimensional conformal field theory, the physics (scattering,
particle production) of the black hole background was further 
analysed in \cite{Dijkgraaf:1991ba}. 

We believe it is important to further develop the analysis of string
theory on black hole backgrounds, starting from first principles. 
Indeed, since string theory is a consistent, unitary theory of quantum
gravity, it should be able to shed light, rigorously, on fundamental
problems like black hole entropy and the black hole information 
paradox. Significant progress was made on both fronts
(see e.g. \cite{Strominger:1996sh}, and \cite{Lunin:2002qf}),
mostly using our improved understanding of 
non-perturbative D-brane physics and supersymmetry.
But we still lack a universal insight into the 
microstates that make up the black hole entropy, and a clear
resolution of the information paradox.

Thus it remains important to analyse simple black hole backgrounds
in string theory, and to uncover more of the quantum gravitational
stringy aspects of singular backgrounds. The $AdS_3$ (or BTZ) black hole
\cite{Banados:1992wn}\cite{Banados:1993gq}
presents itself as a good candidate for further study. Firstly,
it arises in string theory from a near-horizon limit of (for example)
an $F1-NS5-\mbox{momentum}$ 
system. The three-dimensional black hole is therefore a factor
of a spacetime with consistent string propagation. Moreover,
the BTZ black hole is locally $AdS_3$. String propagation on 
$AdS_3$ backgrounds has been analysed in great detail, in the wake
of the $AdS/CFT$ correspondence 
(see e.g. \cite{Giveon:1998ns}\cite{deBoer:1998pp}\cite{Kutasov:1999xu}).
Even more importantly, the detailed and 
rigorous analysis of $Sl(2,R)$ and $Sl(2,C)/SU(2)$
conformal field theory in \cite{Teschner:1999ft}
and 
\cite{Maldacena:2001hw}\cite{Maldacena:2001kv}\cite{Maldacena:2001km} has
provided us with an entirely consistent picture of string propagation on
$AdS_3$ backgrounds, and some more insight 
into the dual conformal field theory.
Now, since the BTZ black hole differs from $AdS_3$ by global identifications
only \cite{Banados:1993gq}, we can view string propagation on BTZ black holes 
as a discrete orbifold of string propagation on $AdS_3$
\cite{Horowitz:1993wt}\cite{Kaloper:1993kj}. Thus, in principle, it seems
 that one can rigorously address fundamental questions on black
hole physics in string theory using these backgrounds. 
More ground needs to be covered though, before a detailed analysis of the
lorentzian $Sl(2,R)$ or euclidean $Sl(2,C)/SU(2)$ orbifold CFT is feasible. 
We take a few first, systematic, steps towards this goal.

The plan of our paper is the following.  In section \ref{space},
we review the backgrounds that we are interested in, i.e.
the space of BTZ black holes. Then we study the dynamics of classical 
particles in BTZ
backgrounds in section \ref{geodesics}, and make a connection to the
classical dynamics of strings.
Next we concentrate on scalar fields and show how
the analysis on a generic BTZ black hole simplifies when an extremal limit
is taken. This nicely fits into special function theory and will allow
us to connect to $Sl(2,R)$ representation theory later.
We then go on to link the
global properties of BTZ black hole backgrounds with preferred parametrisations
of the $Sl(2,R)$ group manifold (and comment briefly on a recent 
cosmological application).
In section \ref{massless}, we 
concentrate  on the massless black hole background and analyse
a few aspects of the string worldsheet CFT in this background. In particular,
we are able to construct the boundary spacetime conformal algebra in terms
of operators on the string worldsheet (in a free field approximation),
 and we show the connection between winding
strings in the vacuum black hole background and a proposal for a ghost 
free spectrum for strings on $Sl(2,R)$.
We conclude and point out more open problems in section \ref{conclusions}.

\section{BTZ black holes}
\label{space}
The low-energy description of closed superstrings  is supergravity.
The supergravity equations of motion have solutions
that include a metric factor that is the BTZ black hole. These solutions
 can be obtained,
for instance,  by taking the near-horizon limit of (macroscopic) 
fundamental strings, wrapping an $S^1$ of a $T^5$,
NS5-branes that wrap the torus, and momentum modes along the fundamental
strings\cite{Hyun:1997jv}\cite{Horowitz:1996ay}\cite{Cvetic:1996xz}\cite{Maldacena:1998bw} 
(or a U-dual thereof).
 Our main interest is in the non-trivial physics associated to the
factor in the metric that corresponds to a BTZ
black hole \cite{Banados:1992wn}. We therefore 
briefly review the space of BTZ black hole metrics.

The BTZ black holes in three-dimensional 
anti-de Sitter space are parametrized by their mass $M$ and angular momentum
$J$. The metric and global identification are given by:
\begin{eqnarray}
ds^2 &=& -(-M+\frac{r^2}{l^2}+ \frac{J^2}{4 r^2}) dt^2
+ (-M+\frac{r^2}{l^2}+ \frac{J^2}{4 r^2})^{-1} dr^2
\nonumber \\ & &
+ r^2 (d \phi - \frac{J}{2 r^2} dt)^2
\label{genmet}
\end{eqnarray}
where $\phi \in {[}0, 2 \pi {[}$.
The mass $M$ and angular momentum $J$ of the black hole are expressed
in terms of the outer and inner horizon as 
$M = \frac{r^2_+ + r^2_-}{l^2}$ and $J= \pm \frac{2 r_+ r_-}{l}$.
The inner and outer horizon are then located at the positive
square root of:
\begin{eqnarray}
r^2_{\pm} &=& \frac{M l^2}{2} (1 \pm (1-(\frac{J}{Ml})^2)^{\frac{1}{2}}).
\end{eqnarray}
The all-important global properties of the BTZ black holes were 
analysed in detail in
\cite{Banados:1993gq}.\footnote{We note moreover that
 we will always suppose
that the background NSNS twoform $B$
is such as to complete the
2d WZW conformal field theory  ($B=\frac{1}{2} r^2 d\phi \wedge dt$) 
on the string worldsheet\cite{Horowitz:1993wt}.}

\subsection*{Supersymmetry}
\begin{figure} 
 \epsfxsize=10cm
\epsfbox{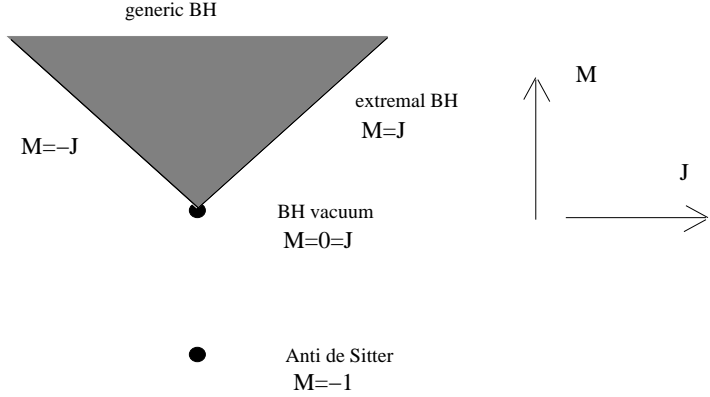}
\caption{\em The space of BTZ black holes \em \label{btzspace}}
\end{figure}

In string theory, we are often interested in backgrounds that preserve
some supersymmetry, for stability and simplicity. The supersymmetric
BTZ black holes  \cite{Coussaert:1994jp}
are the extremal ones. 
These are the
BTZ vacuum black hole with $M=0=J$, and the extremal black hole with mass
$M=|J|$ different from zero. Later, we will focus  on the
vacuum BH. We note that anti-de Sitter space is
formally obtained as the BTZ black hole with $M=-1$ and $J=0$.
These facts yield figure \ref{btzspace},  depicting the space of 
BTZ black holes.

\section{Geodesics}
\label{geodesics}
As a preliminary to studying classical strings in the BTZ black hole
background, we study the geodesic structure of the black hole.
The geodesics have been studied in detail before, in
\cite{Farina:1993xw}\cite{Cruz:1994ir}. Our motivation to revisit
the problem is twofold. First of all, we want to extend the 
study in the literature to include spacelike geodesics. On the
one hand, that makes sense because of the fact that negative 
mass excitations that do not violate the Breitenlohner-Freedman
bound \cite{Breitenlohner:bm}\cite{Breitenlohner:jf}
 do not give rise to instabilities in asymptotically
$AdS$ spactimes. On the other hand, spacelike geodesics proved to
be a good starting point for finding long, winding strings that
turn out to have positive mass. A second reason for revisiting the
nice analysis in  \cite{Cruz:1994ir} is the
particular rescaling of variables that is used, which is not 
appropriate for describing the geodesics of the
 supersymmetric vacuum black hole.

We briefly review timelike and lightlike geodesics \cite{Cruz:1994ir} and 
study spacelike geodesics in the BTZ background. 
We define the 
conserved energy and angular momentum (associated to the two Killing vectors 
($\partial_t$ and $\partial_{\phi}$) of the metric 
(\ref{genmet})) as \cite{Cruz:1994ir}:
\begin{eqnarray}
E &=& (-M + \frac{r^2}{l^2}) \frac{dt}{d \tau} 
+ \frac{J}{2} \frac{d\phi}{d \tau} \nonumber \\
L &=& r^2  \frac{d\phi}{d \tau} - \frac{J}{2} \frac{dt}{d \tau}.
\end{eqnarray}
In terms of these conserved quantities, 
we can write the geodesic equations as:
\begin{eqnarray}
\dot{t} &=& l^2 \frac{E r^2-\frac{JL}{2}}{(r^2-r_+^2)(r^2-r_-^2)}
\nonumber \\
\dot{\phi} &=& l^2 \frac{(\frac{r^2}{l^2}-M)L
+ \frac{JE}{2}}{(r^2-r_+^2)(r^2-r_-^2)} 
\nonumber \\
r^2 \dot{r}^2 &=& -m^2 (\frac{r^4}{l^2}-Mr^2+\frac{J^2}{4})
 \nonumber \\
& & +(E^2-\frac{L^2}{l^2})r^2 + L^2 M - JEL \nonumber
\end{eqnarray}
where $\tau$ is the eigentime, and
$m$ is the mass of the particle. For convenience, we furthermore
define the quantities:
\begin{eqnarray}
\alpha &=& E^2 - \frac{L^2}{l^2} - M m^2 \nonumber \\
\beta &=& L^2 M - \frac{1}{4} m^2 J^2 - JEL. \nonumber \\
\end{eqnarray}
We put $l=1$ from now on, and define a coordinate $y=r^2$ in terms
of which the radial equation simplifies. Then we can 
rewrite the radial differential equation as:
\begin{eqnarray}
\dot{y}^2 &=&  4 m^2(-y^2 +  \frac{\alpha}{m^2} y +  \frac{\beta}{m^2}). 
\label{radeq}
\end{eqnarray}
We have $m^2 > 0$ for timelike geodesics, $m^2=0$ for lightlike
geodesics and $m^2 < 0$ for tachyons.
In the following we solve the radial equation
(\ref{radeq}) for the three different cases. (For further straightforward
integration of the angular and time equation see \cite{Cruz:1994ir}.)
\subsection{Massless particles}
For lightlike geodesics the equations simplify:
\begin{eqnarray}
\dot{y}^2 &=& 4 \alpha y + 4 \beta \nonumber \\
\alpha &=& E^2 - L^2 \nonumber \\
\beta &=& L^2 M- JEL.
\end{eqnarray}
Firstly,
when $E^2 \neq L^2$ we get:
\begin{eqnarray}
y &=& (E^2-L^2) (\tau -\tau_0)^2 - \frac{L^2 M -JEL}{E^2-L^2}.
\end{eqnarray}
Only for $E^2 > L^2$ does the geodesic reach infinity. When
$E^2 < L^2$, the geodesic has a finite range, and reaches a
maximum at $y =\frac{L^2 M -JEL}{L^2-E^2}$.
When $E^2=L^2$, we obtain:
\begin{eqnarray}
y &=& 2 \sqrt{L^2 M- JEL} (\tau - \tau_0).
\end{eqnarray}
Since $E= \pm L$, we have that $L^2 M -JEL \ge 0$.
Thus, 
the geodesic can stretch from infinity to the black hole.
In summary, a massless particle can escape the black hole when
it has sufficient energy \cite{Cruz:1994ir}. 

When $E^2=L^2$ and $L^2M=JEL$ (i.e. $J=\pm M$), we encounter
a special phenomenon (associated to supersymmetry). 
We have the solution
\begin{eqnarray}
y = y_0.
\end{eqnarray}
The lightray then stays at fixed and arbitrary distance 
from the black hole \cite{Cruz:1994ir}.
\subsection{Timelike geodesics}
One can prove that $\alpha^2 + 4 m^2 \beta \ge 0$ for
timelike geodesics (using $M \ge |J|$). The solution to the
radial equation for timelike geodesics is:
\begin{eqnarray}
y &=& \frac{\alpha}{2m^2} + \frac{\sqrt{\alpha^2+4m^2 \beta}}{2m^2}
\sin 2m(\tau-\tau_0).
\end{eqnarray}
Only when $M=\pm J$ and $E=\pm L$ do we obtain the special case
 $\alpha^2+4 m^2 \beta=0$.
Then the massive particle resides precisely
at the horizon of the black hole. Otherwise,
the geodesic stretches over a finite interval in the radial
direction. The particle crosses the outer horizon at some point,
and hits the singularity when $\beta<0$. Otherwise, it oscillates
from inside the inner horizon to outside the outer horizon.
\subsection{Spacelike geodesics}
Here we depart from a review of \cite{Cruz:1994ir}. 
We distinguish two main cases for spacelike geodesics. When
$-4 m^2 \beta > \alpha^2 $ we obtain the solution:
\begin{eqnarray}
y &=& \frac{\alpha}{2 m^2} +\sqrt{-\frac{1}{4 m^4} (\alpha^2+4 m^2 \beta)}
\sinh \sqrt{-4m^2} (\tau-\tau_0).
\end{eqnarray}
The geodesic can stretch from infinity, and will hit the black hole
singularity.
When $\alpha^2+4 m^2 \beta > 0$,
we obtain:
\begin{eqnarray}
y &=& \frac{\alpha}{2 m^2} +\sqrt{\frac{1}{4 m^4} (\alpha^2+4 m^2 \beta)}
\cosh \sqrt{-4m^2} (\tau-\tau_0).
\end{eqnarray}
Now the geodesic has a radius of nearest approach.

The distinction between the two cases is associated to critical
masses (or energy and angular momentum), for a fixed background
($M$ and $J$). We have that
$\alpha^2+4 m^2 \beta={[}(E+L)^2 + (M+J) m^2{]}{[} (E-L)^2 + (M-J) m^2
{]}$. Thus, for fixed $M, J$ and $m^2$, there is a critical 
difference between energy and
angular momentum for which the second solution applies. Then the solution
has  a minimal radius $y_{min}=\frac{\alpha}{2m^2}+
\sqrt{\frac{\alpha^2+4 m^2 \beta}{4 m^2}}$. We will discuss this type of 
solution in a bit more detail for a particular case later.

 In the critical case,
we have either
$ (E+L)^2 + (M+J) m^2 = 0 $ or$ (E-L)^2 + (M-J) m^2 = 0 $.
Then the solution to the radial equation is $y = 
\frac{\alpha}{2 m^2} + c e^{\sqrt{-4m^2} (\tau-\tau_0)}$,
which shows either a stationary solution, or one that stretches 
to infinity.

\subsection*{A special case: the vacuum black hole}
It is no problem to specialize our formulas to the case of the
vacuum black hole (with $M=0=J$), on which we concentrate
later. We briefly summarize the results.

The null geodesics have $\beta=0$. Only when $E^2 \ge L^2$ do we
have a geodesic solution. When $E^2=L^2$ it is the radially stationary
solution.  When $E^2 > L^2$ the solution is given by $r=\sqrt{\alpha}
(\tau-\tau_o)$.

For timelike geodesics we have for $E^2 > L^2$
that $r = \sqrt{\frac{E^2-L^2}{m^2}} \cos m (\tau-\tau_0)$, namely
a particle oscillating near the singularity, and for
$E^2=L^2$ there is only the solution where the particle
sits at $y=0$. 

For spacelike geodesics we find $r = \sqrt{\frac{L^2-E^2}{-m^2}}
\cosh \sqrt{-m^2} (\tau-\tau_0)$ for $E^2<L^2$ which is a solution
with a minimal radius, and 
$r = \sqrt{\frac{E^2-L^2}{-m^2}}
\sinh \sqrt{-m^2} (\tau-\tau_0)$ for $E^2>L^2$, which hits the radial origin
at $\tau=\tau_0$.

\subsection{Discussion}
We gained two things from the extension of the nice
analysis in \cite{Cruz:1994ir}.
On the one hand, we have used formulas that are easily used in the
case of the vacuum black hole. That is a fairly trivial extension.
On the other hand, we have seen that when a tachyonic particle has
sufficient angular momentum, it can stay out of the grasp of the black hole.
That leads us to suspect that there may exist long strings that are
ordinary massive strings, that stay out of reach of the black hole. 
Indeed, when we compare to $AdS_3$\cite{Maldacena:2001hw},
this happens because the long string never shrinks to zero size when it
carries angular momentum. It will remain fat when it is spinning
sufficiently fast. Thus  it may be able to 
keep away from the event horizon of the black hole. 
The special case of the vacuum black hole and the distinction between
the cases where the energy is larger or smaller than the angular
momentum will resurface in later sections.
\section{Scalar fields}
\label{funct}
In this section we review the analysis of the spectrum for scalar
fields in the BTZ black hole background
(see e.g. \cite{Ghoroku:1994ij}\cite{Natsuume:1996ij}\cite{Birmingham:2001dt}).
We will not
push the spectral analysis to the end to review Hawking radiation, quasinormal
modes, etc. Our motivations to include
the abbreviated review are the following. Firstly, we point out how 
to obtain the eigenfunctions of the Laplacian for
the extremal black hole and vacuum black hole background
from the general black hole background by a limiting procedure. 
This unifies some
analyses in the literature and lays bare a technical connection
to results in special function theory which may turn out 
to be useful. Secondly, our analysis will give us the 
opportunity to strengthen the connection of these backgrounds
to $SL(2,R)$ group theory later. 

Indeed, 
the global structure of the generic BTZ black hole differs from the 
global structure of the extremal or 
the vacuum BTZ black hole \cite{Banados:1993gq}\cite{Steif:1996zm}. 
That fact reflects in a preferred  parametrisation
of the $Sl(2,R)$ group manifold when we will try to formulate the 
CFT on the BTZ black hole as an $Sl(2,R)$ orbifold 
(see e.g. \cite{Steif:1996zm}). That choice of parametrisation in turn
reflects a convenient choice of basis for $Sl(2,R)$ representations,
which then give rise to different expression for the matrix elements
(i.e. kernels)
of the representations in terms of special functions. We  find
a foreshadowing of these facts in the following functional analysis,
and will strengthen the connection in section \ref{genrem}.

\subsection{General case}
For a scalar field in a generic BTZ black hole background, 
the Klein-Gordon equation 
$\Delta \Phi = m^2 \Phi(r, \phi,t)$ becomes, after the separation
of variables $\Phi=R(r) e^{-iEt} e^{iL\phi}$:
\begin{eqnarray}
\partial_r \partial_r R + f^{-2} \frac{1}{r} \partial_r (r f^2)
\partial_r R + & & \nonumber \\ f^{-4} (E^2-\frac{JEL}{r^2} + 
(\frac{M}{r^2}-\frac{1}{l^2})L^2) R &=& 4 f^{-2} m^2 R,
\end{eqnarray} 
where $f^2=r^2-M+\frac{J^2}{4r^2}$.
It is convenient to
define a new variable $z = \frac{r_+^2-r_-^2}{r^2-r_-^2}$
and a function $R=(1-z)^{\alpha} g(z)$
in terms of which the differential equation reads:
\begin{eqnarray}
(z(1-z) \partial_z \partial_z -(2 \alpha+1) z \partial_z +
A_1 + B_1 ) g &=& \frac{m^2}{z} g
\end{eqnarray}
where we made use of the quantities:
\begin{eqnarray}
A_1 &=&  
\frac{1}{4} (\frac{E r^+-  L r_-}{r^2_+ - r_-^2})^2 \nonumber \\
    &=& k_+^2 \nonumber \\ 
    &=& - \alpha^2 \nonumber \\ 
B_1 &=& \frac{1}{4}
(\frac{E r^- -  L r_+}{r^2_+ - r_-^2})^2 \nonumber \\
    &=& -k_-^2.
\end{eqnarray}
 Next, by standard manipulations, we obtain the usual form of the 
differential equation for hypergeometric functions. Indeed,
 defining
$g=z^{\beta} X$, $\alpha=i k_+$ and choosing $\beta (\beta-1)=m^2$, 
we obtain:
\begin{eqnarray}
(z(1-z) \partial_z \partial_z +(2 \beta -(2 \alpha+2 \beta +1) z) \partial_z +
& & \nonumber \\
(-\beta^2-2 \alpha \beta +A_1 + B_1) ) X &=& 0.
\label{hyperdif}
\end{eqnarray} 
In terms of the (standard)
parameters 
\begin{eqnarray}
a+b &=& 2 (\alpha+\beta) \nonumber \\
ab &=&
\beta^2+2 \alpha \beta -(A_1 + B_1) 
\nonumber \\
c &=& 2 \beta,
\end{eqnarray} the
solutions to the hypergeometric equations are $F(a,b,c,z)$ and
$z^{1-c} F(a-c+1, b-c+1,2-c,z)$. 
We will choose the parameters $a$ and $b$ to be 
\begin{eqnarray}
a &=&\alpha-ik_- +\beta=
\frac{i(E+ L)}{2 (r_+ + r_-)}+\beta \nonumber
\\
b &=& \alpha+ik_++\beta=
\frac{i(E- L)}{2 (r_+ - r_-)}+\beta.
\nonumber 
\end{eqnarray}  
As usual,
the physical problem under consideration, whether it be Hawking radiation,
quasinormal modes or otherwise, determines the relevant combination
of solutions to the differential equations, by fixing the boundary
conditions. See e.g. \cite{Ghoroku:1994ij}\cite{Natsuume:1996ij}\cite{Birmingham:2001dt}.

\subsection{Extremal limits}

As advertised, the differing global structure of the extremal and massless
BTZ black hole are reflected in the analysis of the spectrum of a
scalar field in these backgrounds. We briefly show how this connects to
limiting procedures in the theory of special functions.

\subsubsection*{Extremal black hole}
Firstly, we concentrate on the extremal black hole with non-zero mass.
The inner and outer horizon coincide for this black hole. We can 
approach the extremal black hole within the BTZ configuration space
by letting the outer horizon $r_+$ tend to the inner horizon $r_-$ from
above, keeping all other quantities fixed. The
 parameter $b=\frac{i(E- L)}{2 (r_+ - r_-)}+\beta$ of
the hypergeometric function becomes
infinite in this limit, the $z = \frac{r_+^2-r_-^2}{r^2-r_-^2}$-coordinate 
tends to zero, while $y=z b$ remains finite. When implementing this
in the differential equation (\ref{hyperdif})
for the hypergeometric function ,
we obtain the equation for the confluent hypergeometric function
\cite{Bateman}:
\begin{eqnarray}
y \partial_y \partial_y S + (c-y) \partial_y S - a S &=&0.
\label{coneq}
\end{eqnarray}
The solutions can be written in terms of Whittaker's functions
\cite{Bateman}.
\subsubsection*{Vacuum black hole}
To obtain the vacuum black hole, we need to take a further limit, along
the space of supersymmetric black holes. When $r_+$ tends to zero
(along a line of supersymmetric black hole backgrounds) we have that 
the parameter $a=
\frac{i(E+ L)}{4 r_+}+\beta$
tends to infinity in (\ref{coneq}). After defining the variable
$x=a y$ which remains finite in this limit,
we obtain the radial equation for the vacuum black hole:
 \begin{eqnarray}
x \partial_x \partial_x T + c \partial_x T -  T &=&0.
\label{besseleq}
\end{eqnarray}
After further redefining $T=x^{\frac{1}{2}(1-c)} U$ and
$w= 2 x^{\frac{1}{2}}$ we find the standard form of the Bessel
equation:
\begin{eqnarray}
w^2 \partial_w \partial_w U+ w \partial_w U-(w^2+(1-c)^2)U &=&0.
\end{eqnarray}
Thus
 the scalar
wave function can be written in terms of Bessel functions with
index $\nu=2 \tau +1$ (where $\tau=-\beta$).

For this case,  which will be of particular interest to us later,
we give a little more detail. In terms of the
original $(r,\phi,t)$ coordinates, we have, for $E^2>L^2$:
\begin{eqnarray}
\Phi=e^{-iEt} e^{iL\phi} \frac{1}{r} (c_1 J_{2 \tau +1}
(\frac{\sqrt{E^2-L^2}}{r})+c_2 J_{-(2 \tau +1)}
(\frac{\sqrt{E^2-L^2}}{r}))
\end{eqnarray}
and the solution regular in the interior for $E^2<L^2$ is:
\begin{eqnarray}
\Phi=c e^{-iEt} e^{iL\phi} \frac{1}{r}  K_{2 \tau +1}
(\frac{\sqrt{L^2-E^2}}{r}).
\end{eqnarray}
Without going into the details of important special cases (for instance
discussed in the context of $AdS_3$ in 
\cite{Balasubramanian:1998sn}\cite{Kutasov:1999xu}),
we remark that for $\tau$ real, we have for $E^2>L^2$ an oscillating
solution that is localized near the origin. For $2 \tau+1$
purely imaginary (i.e. $m^2 \le -\frac{1}{4}$),
 we find solutions that travel to
the boundary. In the timelike case, we find incoming and outgoing waves,
while in the second case these waves 
are related by demanding regularity. Thus,
the picture that emerges is consistent with the analysis of geodesics
presented at the end of section \ref{geodesics}, where we found
timelike geodesics concentrated near the singularity, and
spacelike geodesics that could travel to the boundary.\footnote{Our choice 
of values
for $\tau$ will find justification in the next section, from the
representation theory of $Sl(2,R)$.} 
\subsection{Summary}
The limiting procedures in special function theory that relate the
hypergeometric functions to confluent hypergeometric functions
and  Bessel functions
have a natural interpretation as limiting procedures for scalar wave
functions in generic, extremal, or massless BTZ black hole backgrounds.
This is a foreshadowing of the global picture we paint in the next
section.

\section{Strings and BTZ black holes}
\label{genrem}
In this section we will assemble some general remarks on strings
in BTZ black hole backgrounds. We stress the importance of the 
group theoretic approach, which should enable us to make use of results
of WZW-models, generalized to non-compact groups. We
 recall some facts on the analysis of strings in
$AdS_3$ backgrounds, and their connection to $Sl(2,R)$ representation
theory. Next, we discuss the different group parametrisations suited
for a generic, an extremal and a vacuum BTZ black hole background,
and connect these parametrisations to the special function theory of
the previous section. Then we point out the generic occurence of wound
strings in these backgrounds, classically. We point out the close
analogy of the generic BTZ black hole to a recently studied stringy
cosmological background, and point out a dilemma.

This section could serve as a generic framework for discussing strings
in BTZ backgrounds algebraically. In the next section, we will restrict
attention to the case that interests us most in this paper,
 the vacuum black hole.

\subsection{Strings on $AdS_3$}
In this subsection, we review some features of strings on $AdS_3$ that
will find a generalisation in strings in BTZ backgrounds. We follow
the systematic treatment of
\cite{Maldacena:2001hw} 
to which we the reader for a lot more details.
The manifold $AdS_3$ (with no closed timelike loops) is
the covering space of the  group manifold $Sl(2,R)$. To study
strings in $AdS_3$, it is useful to study the $Sl(2,R)$ Wess-Zumino-Witten
model, where we parametrize the group elements as
\begin{eqnarray}
g=e^{ \frac{i}{2}(t + \phi)\sigma^2} e^{\rho \sigma^3} e^{
\frac{i}{2}(t-\phi) \sigma^2}. \label{ads} 
\end{eqnarray}
One needs to be careful in defining the configuration space,
though, to ensure that the strings do live on the covering of the
group manifold $Sl(2,R)$ --  we unwrap the time coordinate $t$.

A strong motivation for the choice of parametrisation
(\ref{ads}) is that it allows for a straightforward diagonalisation
of the energy and angular momentum, by a choice of basis in
$Sl(2,R)$ representations that diagonalize $J_0^3$ and $\bar{J}_0^3$,
which generate time translation and rotations
\cite{Maldacena:2001hw}. Thus, by our choice
of parametrisation, it is easy to diagonalize
the action of an elliptic subgroup of $Sl(2,R)$ (see \cite{VK}
p.377).

Excitations in $AdS_3$ string theory fall into
representations of (the affine extension of)
the covering group of $Sl(2,R)$. 
The unitary representations
of the covering are enumerated in the appendix. The demand of unitarity 
yields a first restriction on  the possible masses for particles
in $AdS_3$ spaces. These restrictions 
consequently lead to a restriction on the
string Hilbert space.

Within the set of unitary representations of the covering
group of $Sl(2,R)$,  we further
restrict ourselves  to two
types of unitary irreducible representations, namely the principle
discrete and the principle continuous representations. (We thereby 
disregard the principle complementary series, which do not
arise as quadratically 
integrable particle wavefunctions on $AdS_3$.)

Due to the $\widetilde{(Sl(2,R)} \times \widetilde{Sl(2,R)})/Z_2$ 
background isometrygroup, we know
the particle excitations will form representations of $\widetilde{Sl(2,R)}$.
More precisely, the particle wavefunctions will be tensor
representations of the $\widetilde{(Sl(2,R)} \times \widetilde{Sl(2,R)})/Z_2$ 
symmetry-group,
which can  be represented as matrix elements of the $\widetilde{Sl(2,R)}$
representation $R_{\chi}$ labelled by $\chi=(\tau,\epsilon)$, with quadratic
Casimir $c_2 = -\tau(\tau+1)$ \cite{VK}\footnote{The parameter
$\epsilon \in {\{} 0,\frac{1}{2} {\}}$ labels whether the center
is represented non-trivially. It will not play a crucial role
in the following.}. 
The mass squared
of the particle is $m^2=\tau(\tau+1)$.

The principle continuous series of representations has $\tau=-\frac{1}{2}
+is$ such that $m^2= - \frac{1}{4} - s^2$, in violation of the 
Breitenlohner-Freedman bound. This indicates that these are unstable modes
(in a spacetime that is asymptotically $AdS$). The representations are
associated to spacelike geodesics in the classical picture. 
We moreover have the principle discrete series with 
$m^2 \ge  - \frac{1}{4}$, such that (most of) these correspond to 
ordinary particles in $AdS_3$.

\subsection*{Spectral flow}
It was further argued in 
\cite{Maldacena:2001hw}\cite{Maldacena:2001kv}\cite{Maldacena:2001km} that
the solutions obtained by acting with  a non-trivial automorphism on
a given geodesics solution to the equations of motion, yields winding strings
\cite{Maldacena:1998uz}\cite{Seiberg:1999xz}
that are crucial for the determination of the spectrum for strings on 
$AdS_3$ backgrounds. We refer to \cite{Maldacena:2001hw} for an
extensive discussion.

\subsection{Black Hole}

Similarly, we want to make some exploratory remarks on
 string theory on BTZ black hole backgrounds.
We will show that different regions of spacetime
require different parametrisations of the $Sl(2,R)$ group elements.
We link this picture to recent explorations of cosmological
models in string theory in passing.

\subsubsection*{Generic Black Hole}
For the case of the generic black hole, we assemble a few perhaps well-known
facts.
First of all we note that we can write every $Sl(2,R)$ group element 
with all non-zero matrix elements as
follows\footnote{A separate 
treatment of the group elements with a zero entry
is necessary.}: 
\begin{eqnarray}
g & = & e^{ \frac{u}{2} \sigma^3} (-1)^{\epsilon_1} .(i \sigma^2)^{\epsilon_2}
.p.
e^{\frac{v}{2} \sigma^3} \label{gen}
\end{eqnarray}
where $\epsilon_{1,2} \in {\{} 0,1 {\}}$ and
the two by two matrix $p$ takes one of the following two forms:
\begin{eqnarray}
p_1 &=& e^{ \rho \sigma_1} \\
p_2 &=& e^{ i \rho \sigma_2}.
\end{eqnarray}
Next we note that 
we can divide the $Sl(2,R)$ group manifold into 3 large regions (each
consisting of 4 smaller regions). (See e.g. \cite{Elitzur:2002rt} for 
a discussion in a physical context.) 
When we take the convention that 
$g= \left( \begin{array}{cc} a & b \\
                             c & d 
           \end{array} \right) $, 
the three regions are characterized
by the signs of $ad$ and $bc$. Region $I$, where $ad>0$ and $ bc>0$ 
can be shown  \cite{Banados:1993gq} to correspond to the region 
outside the outer horizon, region $II$  
($ad>0$ and $ bc<0$) to the region between the inner
and the outer horizon, and region $III$ ($ad<0$ and $ bc<0$) 
to a region inside the inner
horizon. Each of the twelve regions can be parametrized in terms of a
group element written as in (\ref{gen}).
The precise correspondence can be found  in \cite{Elitzur:2002rt}.

As an example, consider part of 
the region $ad>0$ and
$bc>0$. We can parametrize the $Sl(2,R)$ group element as:
$g=e^{ \frac{u}{2} \sigma_3} 
p_1 e^{\frac{v}{2} \sigma_3}$. From this we can easily find
the metric, and we can transform to the usual Schwarschild type coordinates
 via the coordinate transformation:
\begin{eqnarray}
\cosh^2 \rho &=& \frac{r^2-r_-^2}{r_+^2-r_-^2} \nonumber \\
u &=& (r_+ - r_-) (t+\phi) \nonumber \\
v &=& (r_+ + r_-) (\phi-t). 
\end{eqnarray}
Clearly, this describes a region of spacetime beyond the outer
horizon.
An important point is that the parametrisation (\ref{gen}) is suited
for the diagonalisation of a hyperbolic subgroup of $Sl(2,R)$ 
\cite{Natsuume:1996ij}\cite{Hemming:2001we}.
For this background the hyperbolic parametrisation naturally 
leads 
to diagonalisation of the energy and angular momentum, since
these are associated to hyperbolic generators in this background.
The matrix elements in that basis can be written down in terms
of hypergeometric functions (\cite{VK} section 7.2).
 This squares nicely with the functional analysis
in section \ref{funct} and shows that an algebraic analysis
close to the two-dimensional treatment in \cite{Dijkgraaf:1991ba}
of black hole scattering, Hawking radiation, etc, can be
repeated in this context.

\subsubsection*{Brief remark}
We point out that there is some unresolved tension between two 
reasonable points of view in the literature. 
To avoid closed timelike curves, the authors of \cite{Banados:1993gq}
cut out part of region III from the spacetime, and it was
moreover argued in a general relativistic context
that matter couplings (and a curvature singularity
that could arise from matter at $r=0$) would force one to do so
(see  \cite{Banados:1993gq} section V).

In the stringy and algebraic context of
a coset WZW conformal field theory the authors of 
\cite{Elitzur:2002rt} kept the region III  
(compare figure 2, region III in \cite{Banados:1993gq} 
to figure  1  in \cite{Elitzur:2002rt})
 which seemed natural from a group theoretic perspective. 
They pointed out that the physics of closed timelike curves needs
further study in their stringy cosmological backgrounds. 

Although the contexts of application of the $Sl(2,R)$ group theory
are different, it seems to us that there is some tension between
these two approaches. We will not try to resolve it here, but believe
it is crucial to compare these backgrounds in more detail
on this particular point (and decide on whether it makes sense
to add the regions of spacetime that contain closed timelike curves
within a string theoretic context).

\subsection{Extremal black holes}
A similar story can be told for extremal black holes (see also
\cite{Maldacena:1998bw}\cite{Satoh:1998sg}),
where a mixed basis parametrisation
leads to straightforward diagonalisation of an parabolic
and a hyperbolic subgroup, corresponding to energy and angular
momentum generators.
The kernels can in this case be written in terms of
Whittaker functions (\cite{VK} section 7.7.3 and 3.5.7). 
For the vacuum black hole, two parabolic subgroups are to be
diagonalised on physical grounds. Then the kernels (i.e. matrix
elements) may be written in terms of Bessel functions (\cite{VK}
section 7.6) in full
agreement with the analysis in section \ref{funct}.

\subsection{Winding strings}
\label{gensol}
After stressing the physical logic (i.e. diagonalisation of conserved
quantities) behind the global parametrisations
of the $Sl(2,R)$ group manifold, it becomes straightforward to discuss
the particular classical solutions corresponding to strings winding
the BTZ black hole.
The propagation of strings in black hole backgrounds will be described
by an (orbifolded) $Sl(2,R)$ WZW model (where different parametrisations are
appropriate  according to the background and particular patch of
spacetime) (see also \cite{Steif:1996zm}\cite{Hemming:2001we}).
The classical solutions to the string equations of motion
for a WZW model
are  given by a product of left and right movers
as $g= g^+(x^+) g^-(x^-)$ (where $x^{\pm}=\tau \pm \sigma$).
After our preliminary investigations, 
it is easy to state the generalisation of the solution
to the equations of motion that will provide us with the
winding strings in black hole backgrounds. We can start from
a solution to the equations of motion that represents a pointlike
string moving on the geodesics discussed in section \ref{geodesics}.
Then we make use of the following general technique.

The closed string only has to close
up to the action of an element of the orbifold group, which 
implements the periodicity in the angular coordinate $\phi$. 
Thus we have that the string (given by an embedding in the group
manifold) closes  as
$g(\sigma + 2 \pi) = (h_l)^n g^+(x^+ +2 \pi) g^-(x^--2 \pi) (h_r)^n$,
where $h=(h_l,h_r)$ is the generator of the infinite discrete
orbifold group (which is a subgroup of $Sl(2,R)_L \times Sl(2,R)_R$).

A typical solution in the $n$-twisted sector would be $g^+(x^+ + 2 \pi)
= (h_l)^n g^+(x^+)$ and $g^-(x^- - 2 \pi)
=  g^-(x^-) (h_r)^n$ where $n$ labels the twisted
(winding) sectors of the orbifold. The  modification from
the solutions that represent pointlike geodesic motion are easily
found from the appropriate parametrisations of the group elements in
all backgrounds and patches. (This is straightforward because we made
sure that the parametrisations are such that the energy and angular
momentum are naturally diagonal. Formally speaking,
 we made sure that the outer
factors of the group element decomposition always contain the angular
and time coordinate.) Thus, we constructed the classical
winding string solution in all generality. We refrain from enumerating
in great length all cases -- in the next section we will have the
opportunity to study one particular case in detail.

Although the above reasoning is sufficient to establish the existence
of these classical
solutions,  we included an explicit parametrisation and
check of the solutions in  appendix \ref{clasgen}, in an 
alternative formalism.

\section{Vacuum black hole}
Previous sections were concerned with coming to grips with fitting the
mathematical results on $Sl(2,R)$ group theory into the physical framework
of BTZ black holes, while obtaining some more insight into the geodesic
structure of BTZ black holes, the global structure of the spacetime and
of $Sl(2,R)$, and the relation to results in the theory of special 
functions. Apart from the new physics that we uncovered by a careful
review and extension, our analysis also prepared us for some of the
new phenomena we will encounter in a more detailed study of the particular
case of the vacuum black hole.

The metric for the vacuum black hole with $M=0=J$
is given by 
\begin{eqnarray}ds^2 &=& - \frac{r^2}{l^2} dt^2
+ \frac{l^2}{r^2} dr^2 + r^2 d\phi^2
\end{eqnarray}
with $\phi \in {[}0, 2 \pi {[}$ (and we put $l=1$ again
from now on). 
Following the general logic,
we define the group element\footnote{To be a little
more precise, every $Sl(2,R)$ group element can
 be parametrized as follows.
Either as $g= e^{v_l L^-} e^{\rho \sigma^3} 
.(i \sigma^2).e^{-v_r L^-} (-1)^{\nu}$
or, for group elements with a zero: 
$g=e^{v_l L^-} e^{\rho \sigma^3}(-1)^{\nu}$. We mod out by $-1$ (and
put $\nu=0$). The group elements with a zero should 
be thought of as located at the black hole singularity. We do not
discuss the subtleties associated to this patch here.} 
\begin{eqnarray}
g &=& e^{v_l L^-} e^{\rho \sigma^3}.s.  e^{-v_r L^-} 
\end{eqnarray}
where $\sigma^i$ are the Pauli matrices and $L^{\pm}=\frac{1}{2}
(\sigma^1 \pm i \sigma^2)$.
The parametrisation
is related to the previous coordinates by the transformation
\begin{eqnarray}
v_l &=&  \phi +t  \\
v_r &=& \phi - t \\
e^{\rho} &=& r
\end{eqnarray}
In terms of these coordinates the metric reads:
\begin{eqnarray}
ds^2 &=& e^{2 \rho} dv_l dv_r + d \rho^2,
\end{eqnarray}
which is the invariant metric on the group manifold.
The global angular identification in these coordinates is
$(v_l,v_r,\rho) \equiv (v_l+2 \pi, v_r+ 2 \pi,\rho)$,
which is generated by $(h_l,h_r)=(e^{2 \pi L^-},e^{-2 \pi L^-})$.

We can define the currents as $J^i_L =k Tr(L^i \partial_{+}g g^{-1})$
for the leftmovers, and
$J^i_R =k Tr(L^i g^{-1} \partial_{-}g )$ for the rightmovers,
where $L^3=\frac{i}{2} \sigma^3$.
The leftmoving currents can be computed to be:
\begin{eqnarray}
J^3_L &=& ik ( \partial_+ \rho- e^{2 \rho} v_l \partial_+ v_r) \nonumber \\
J^-_L &=& k e^{2 \rho} \partial_+ v_r \nonumber \\
J^+_L &=& k( \partial_+ v_l+ 2 v_l \partial_+ \rho- v_l^2 e^{2 \rho} 
\partial_+ v_r)
\end{eqnarray}
and the rightmoving ones are:
\begin{eqnarray}
J^3_R &=& -ik ( \partial_- \rho- e^{2 \rho} v_r \partial_- v_l) \nonumber \\
J^-_R &=& -k e^{2 \rho} \partial_- v_l \nonumber \\
J^+_R &=& -k (\partial_- v_r+ 2 v_r \partial_- 
\rho - v_r^2 e^{2 \rho} \partial_- v_l),
\end{eqnarray}
where we used the cylindrical 
worldsheet coordinates $x^{\pm}=\tau \pm \sigma$.

\subsection{Winding sectors}
We obtain the twisted sector of the orbifolded theory by generalized
spectral
flow from the untwisted sector. As explained before 
(in subsection \ref{gensol}),
we can add winding to a
classical (e.g. geodesic) solution  $\tilde{g}=\tilde{g}^+
\tilde{g}^-$ by the substitution:
\begin{eqnarray}
g^+ &=& e^{ n x^+ L^-} \tilde{g}^+ \nonumber \\
g^- &=& \tilde{g}^-  e^{ n x^- L^-}.
\end{eqnarray}
Clearly, this generates a new solution to the equations of motion.
Moreover, if the original solution belonged to the untwisted sector,
than the new solution belongs to the sector twisted by $n$ units,
since $g(\sigma+2 \pi)=g^+(x^+ + 2 \pi) g^-(x^- - 2 \pi)
=e^{2 \pi n L^-} g(\sigma) e^{-2 \pi n L^-}$.

The action of this operation on the currents is:
\begin{eqnarray}
J^-_L &=& \tilde{J}^-_L \nonumber \\
J^3_L &=& \tilde{J}^3_L - in x^+ \tilde{J}^-_L \nonumber \\
J^+_L &=&  \tilde{J}^+_L -      (n x^+)^2 \tilde{J}^-_L 
    - 2i n x^+ \tilde{J}^3_L+kn, 
\label{clascur}
\end{eqnarray}
and similarly for the rightmovers.
The action on the energy momentum tensor for right- and leftmovers
is given by:
\begin{eqnarray}
T_{++} &=& \tilde{T}_{++} + n \tilde{J}^-_L + T_{++}^{other}
\nonumber \\
T_{--} &=& \tilde{T}_{--} + n \tilde{J}^-_R + T_{--}^{other}
\label{trt}
\end{eqnarray}
where we used the conventions that $T_{++}^{other}$ corresponds
to the part of the string wordlsheet CFT that describes the
seven directions transverse to the BTZ black hole. (Our
conventions are similar to the ones in  \cite{Maldacena:2001hw}.)
That implies the following transformation rule for the 
worldheet zeromodes of the stress tensor:
\begin{eqnarray}
L_0 &=& \tilde{L_0} + n \tilde{J}^-_{L,0} + L_0^{other}
\nonumber \\
\bar{L_0} &=& \tilde{\bar{L_0}} + n \tilde{J}^-_{R,0} + L_0^{other}.
\end{eqnarray}

Now, the time translation generator and angular momentum 
generator in this background are given by:
\begin{eqnarray}
E &=& -(J^-_{L,0} + J^-_{R,0}) \nonumber \\
L &=& J^-_{L,0} - J^-_{R,0}. \label{clasen}
\end{eqnarray}
Note that the transformation properties of the currents imply that
the energy of the strings in the twisted sectors are degenerate
(but the constraint equations will require different internal
conformal weights).

When we assume  that $\tilde{T}_{++}=-k m^2$ (as for
a geodesic) and
$T_{++}^{other}=h$ for the internal conformal field theory,
 and similarly for the rightmovers, 
the zeromode constraint equations in the twisted sectors read:
\begin{eqnarray}
-k m^2 + \frac{n}{2} (L-E) + h = 0
\nonumber \\
-k m^2 + \frac{n}{2} (-L-E) + \bar{h} = 0.
\label{constr}
\end{eqnarray}
Thus, the internal conformal weights are related to the
spacetime angular momentum by $\bar{h}-h=n L$. The
spacetime angular momentum is quantized, i.e. $L\in Z$.

For $n=0$, the only solutions to the constraints (\ref{constr})
are associated
to timelike geodesics.
(We assume that the conformal weights $h$ and $\bar{h}$
associated to the internal conformal field theory are positive.)
But now we notice that in the twisted
sectors, we can find  solutions to the
constraints that arise from spacelike geodesics for $E^2>L^2$
and $n$ positive. 
The fact that the spacelike geodesics are promoted to viable
solutions to the constraints in the twisted sector
is similar to the picture obtained in \cite{Maldacena:2001hw}.

\section{Quantum CFT description} 
\label{massless}
\subsection{Spectrum and ghosts}
Now we come to an important point. The classical current algebra
(\ref{clascur})
in the twisted sector is the classical counterpart of the current algebra 
that was introduced in \cite{Bars:1995mf} ( p.4 (3.2))\footnote{Note that 
\cite{Bars:1995mf} works with a worldsheet coordinate on the
plane were we worked on the cylinder in the previous section.}  and
\cite{Bars:1995cn}. 
There, a modification of the currents
was formally introduced to
obtain a ghost 
free spectrum for string theory on $Sl(2,R)$ backgrounds.\footnote{A
formal connection to a generic BTZ background was proposed in
\cite{Satoh:1997xe}.}
 We see that these
currents
arise natural in our approach when analysing winding strings in the
vacuum black hole background.

We can then apply the results of \cite{Bars:1995mf}
in this context. Specifically, we will find physical states in our
quantum Hilbert space that satisfy (compare
also to the analogous treatment in $AdS_3$ \cite{Maldacena:2001hw}):
\begin{eqnarray}
L_0-1 |\tilde{\tau}, h, \tilde{N}, j_0^- \rangle &=&
 -\frac{\tilde{\tau}(\tilde{\tau}+1)}{k-2}+ 
\tilde{N}+h-1+n j_0^- |\tilde{\tau}, h, \tilde{N}, j_0^- \rangle 
=0 \nonumber \\
\bar{L}_0-1 |\tilde{\tau}, h, \tilde{\bar{N}}, \bar{j}_0^- \rangle &=& 
-\frac{\tilde{\tau}(\tilde{\tau}+1)}{k-2}+ 
\tilde{\bar{N}}+\bar{h}-1+n \bar{j}_0^-
 |\tilde{\tau}, h, \tilde{\bar{N}}, \bar{j}_0^- \rangle =0,
\end{eqnarray}
where the tilded quantities $\tilde{\tau}$ and $\tilde{N}$
refer to Casimir and oscillation number before
twisting. The quantity $j_0^-$ is the eigenvalue of the
$J_0^-$ operator.
 The quantisation of angular momentum, equivalent to periodicity
in the spacetime coordinate $\phi$, is now the natural interpretation of the
``monodromy'' projection in \cite{Bars:1995mf}. It was proven in \cite{Bars:1995mf} that restricting to continuous representations gives a ghost free
spectrum. We will not analyse in detail how the discrete representations
fit into the picture, although we expect them to occur in the spectrum,
as in the $AdS_3$ background.
One obstruction to a completion of the spectral analysis is the fact that
the twisted sectors are not  easily expressed in terms of a 
representation of the original current algebra, in contrast with the
$AdS_3$ case.

Now, in \cite{Bars:1999ik} there is
a suggestion that the form of the current algebra is linked to winding
strings in an $AdS_3$ background, resulting in some unresolved tension with
the works \cite{Giveon:1998ns}\cite{deBoer:1998pp}\cite{Kutasov:1999xu}.
We resolved 
the tension by noticing that the logarithmic cuts (in terms of a
planar worldsheet coordinate), 
and the winding strings, are properly interpreted as being apart of the
string conformal field theory in the background of the massless
BTZ black hole.

\subsection{Spacetime Virasoro algebra}
A second issue we want to address in the quantum theory, is
the construction of the spacetime Virasoro algebra in terms
of operators in the worldsheet conformal field theory. For
string theory on $AdS_3$ backgrounds, the construction was
obtained in \cite{Giveon:1998ns}\cite{deBoer:1998pp}\cite{Kutasov:1999xu}.
Since the BTZ black holes are also asymptotically $AdS_3$, we expect
on the basis of the pioneering work \cite{Brown:nw} to find a realisation
of the Virasoro algebra acting on the space of physical states as well.
To obtain a  construction of the spacetime Virasoro generators,
we study the conformal field theory in euclidean signature in 
target space and on the worldsheet (parametrized by
the planar coordinate $z$).

Note first of all that
the conformal field theory description of the euclidean $AdS_3$,
is globally given by a lagrangian:
\begin{eqnarray}
L &=& k (\partial \phi \bar{\partial} \phi + e^{2 \phi} \bar{\partial}
\eta \partial \bar{\eta}),
\end{eqnarray}
where $\eta$ takes values on the plane. By contrast, the euclidean
version of the BTZ vacuum black hole is described by the CFT:
\begin{eqnarray}
L &=& k (\partial \rho \bar{\partial} \rho + e^{2 \rho} \bar{\partial}
\gamma \partial \bar{\gamma}),
\end{eqnarray}
where $\gamma \simeq \gamma + 2 \pi$ 
takes values on the cylinder. We want to discuss the latter CFT.

As described for instance in \cite{Giveon:1998ns} in some detail,
one can approximate the above conformal field theory near the boundary
by a free field theory. The lagrangian 
\begin{eqnarray}
L &=&  \partial \rho \bar{\partial} \rho 
- \frac{2}{\alpha_+} R^{(2)} \rho- \beta \bar{\partial} \gamma
- \bar{\beta} \partial \bar{\gamma} - \beta \bar{\beta} 
e^{-\frac{2}{\alpha_+}\rho},
\end{eqnarray}
where $\alpha_+^2=2k-4$,
is equivalent to the previous one, and we can compute correlation
functions that are dominated by the large $\rho$ region by ignoring
the last term. 

In the free field approximation to the conformal field theory,
we can realize the $Sl(2)$ current algebra:
\begin{eqnarray}
J^3 (z) J^{\pm}(w) \simeq \frac{\pm J^{\pm}(w)}{z-w} \nonumber \\
J^3(z) J^3(w) \simeq -\frac{k}{2} \frac{1}{(z-w)^2} \nonumber \\
J^-(z) J^+(w) \simeq \frac{k}{(z-w)^2} + \frac{2 J^3(w)}{z-w}
\end{eqnarray}
using the free fields:
\begin{eqnarray}
J^3 &=& - \beta \gamma + \frac{\alpha_+}{2} \partial \rho \nonumber \\
J^+ &=& - \beta \gamma^2 + \alpha_+ \gamma \partial \rho +k
\partial \gamma
 \nonumber \\
J^- &=& - \beta
\end{eqnarray}
where we use the conventional OPE for the free fields $\rho(z) \rho(w)
\simeq - \log (z-w)$ and $\beta(z) \gamma(w) \simeq -\frac{1}{z-w}$.

Since the black hole is asymptotically $AdS_3$, we expect a realisation
of the Virasoro algebra to act on the space of physical states \cite{Brown:nw}.
The boundary of our space is $R^1 \times S^1$, because of the identification
on $\gamma$.
First of all, we note that the spacetime ${\cal L}_0$ should measure
spacetime energy and angular momentum (see also (\ref{clasen})), 
and is therefore roughly
given by \footnote{
We will restrict to writing formulas for leftmovers only when the rightmovers
behave in close analogy.}:
\begin{eqnarray}
{\cal L}_0 & \sim & \oint dz J^-(z).
\end{eqnarray}
Now, since we expect the conformal weight of the higher generators
to be related to their index, the index is expected to be related to the
$J^-_0$ charge. We
propose to describe the spacetime Virasoro generators in terms
of the free fields as follows:
\begin{eqnarray}
{\cal L}_n &=& -i\oint dz (J^- + \frac{i n \alpha_+}{2} \partial \rho +  
\frac{k}{4} \partial
\gamma) e^{i n \gamma} (z).
\end{eqnarray}
This expression is also inspired by the form of the
classical Virasoro algebra on the cylinder, the periodicity of
$\gamma$, and the conformal dimensions of the
operators in the theory. It can be checked that the coefficients
of the first
and second term have the right ratio to make sure the integrand is primary.
The coefficient of the third term is such that the anomalous term (linear
in $n$) in
the Virasoro algebra works out.
Indeed, the commutator of these 
Virasoro operators can be evaluated using the contour argument,
and the OPE of the free fields that make up the operators ${\cal L}_n$. 
We obtain the 
algebra:
\begin{eqnarray}
{[} {\cal L}_m, {\cal L}_n {]} &=& (m-n) {\cal L}_{m+n} + 
\frac{c}{12} (m^3-m) \delta_{m+n}, 
\end{eqnarray}
where $c=6 k w$ and $w = i \oint \partial \gamma$ measures the winding 
number of the string at infinity. This is a nice black hole counterpart
to the construction presented in \cite{Giveon:1998ns}. It would be
interesting to understand the analogue of this construction for strings
on general BTZ backgrounds.
Note that the Virasoro generator ${\cal L}_n$
 acts on the field $\gamma$ roughly
as ${\cal L}_n \sim e^{in \gamma} \partial_{\gamma}$, as expected from
the standard form of the conformal algebra on the cylinder. 

Moreover, it should be clear that 
if a further affine symmetry (with currents $K^a(z)$) 
is present on the worldsheet, arising
from the internal CFT, we can construct a corresponding target
space affine algebra in a manner formally
similar to \cite{Giveon:1998ns} by introducing the spacetime
operators $T_n^a = \oint dz K^a(z) e^{in \gamma}$. The target
space current algebra will have a level equal to the level
of the worldsheet current algebra, multiplied by the winding
number. 
\subsection*{Remark on wound free fields}
To describe the winding strings in the quantum
picture, we can introduce a new field $\alpha$ such that $\beta \equiv
\partial \alpha$ 
and
\begin{eqnarray}
\alpha(z) \gamma(w) \simeq - \log (z-w).
\end{eqnarray}
We can define the current $J_{\gamma}=-i \partial \gamma$. We expect
the zeromode to describe momentum and winding in the direction 
$\gamma$, which is the direction around which we want to wind the string.
The operator $e^{i q \alpha}$ has charge $q$,
and can therefore be used to introduce vertex operators corresponding
to winding strings.

We will not attempt to give a complete discussion of vertex operators here,
but one should be able to recuperate  results of 
\cite{Bars:1999ik} in this context. It would certainly be interesting
to revisit the discussion in \cite{Bars:1999ik} in 
the light of our conceptual clarification. 

\section{Conclusions and future directions}
\label{conclusions}
In this paper, we analysed string theory on $AdS_3$ black hole
backgrounds. We argued strongly for the importance of an appropriate 
parametrisation of the $Sl(2,R)$ group manifold. Depending
on whether one is studying the massless black hole, the
extremal black hole, or a generic black hole, one will
choose a parametrisation in which two parabolic subgroups
are easily diagonalised, a parabolic and a hyperbolic
subgroup, or two hyperbolic one parameter
subgroups. That choice allows
for an algebraic treatment of the eigenfunctions of the
Laplacian, as well as for a straightforward implementation
of the orbifolding of the WZW theory. In the 
case of the generic BTZ black hole, we pointed out a close
analogy to  recent work on stringy
cosmology \cite{Elitzur:2002rt}, 
and showed the tension between the algebraic
and the general relativistic approach.

By analysing geodesic motion, and classical string solutions,
we gained intuition for the spectrum of strings on BTZ backgrounds.
Next, we focussed on the massless BTZ black hole background, because
of its simplicity, supersymmetry, its connection to pure $AdS_3$,
and its relation to recent attempts to study cosmological backgrounds
in string theory (see e.g. \cite{Elitzur:2002rt}\cite{Simon:2002ma}\cite{Liu:2002ft}\cite{Craps:2002ii}\cite{Lawrence:2002aj} for references closest
to our work).
By taking the global structure of the space properly into account,
and using a classical analysis of wound strings, we argued that the 
proposal for a ghost free spectrum of \cite{Bars:1995cn} finds a natural
interpretation as pertaining to strings on the massless black hole
background.
It follows that the results of \cite{Bars:1999ik} can be reinterpreted 
as applying to the AdS/CFT duality for the massless
BTZ black hole.

As a first important application of our analysis, we provided the
construction of the spacetime Virasoro algebra in the massless
black hole background. Thus, we gave another
 explicit realisation of
the spacetime boundary conformal algebra 
in a quantum theory of gravity with propagating degrees
of freedom.

There are many interesting directions for future research. Of 
particular interest would be the rigorous analysis of the exact spectrum
(along the lines of \cite{Maldacena:2001kv}\cite{Hanany:2002ev}) 
for strings in the massless black hole background,
preferably in a supersymmetric context (see e.g. \cite{Argurio:2000tb}),
 such that, after applying the
$AdS_3/CFT$ correspondence, it can be compared, via spectral flow
in the spacetime boundary conformal field theory,
to the spectrum of strings on $AdS_3$ (see e.g. 
\cite{Lunin:2001jy}\cite{David:2002wn} and references therein). 
That would provide a nice check
of the consistent spectrum for strings on a black hole background.
Other directions are the application of the orbifold procedure
on the euclidean $Sl(2,C)/SU(2)$ CFT so as to describe strings propagating on
the euclidean BTZ black holes \cite{Carlip:1994gc}.

In 
\cite{Maldacena:2001km} the scattering amplitude for two long wound strings 
in $AdS_3$ was suggested
as an interesting computation, because of its interpretation in terms
of an S-matrix in an $AdS_3$ background. It would be even more interesting
if the scattering process would yield information on the production rate
of black holes in a quantum theory of gravity.
In the context of classical
three-dimensional gravity, it was shown
that the collision of two fast particles in $AdS_3$ would yield a black hole
\cite{Matschull:1998rv}.
 (See also e.g. \cite{Balasubramanian:2000rt}\cite{Martinec:2001cf}.)

We hope to return to some of these issues in the future.

\acknowledgments
It is a pleasure to thank my
 colleagues at MIT for stimulating conversations on diverse topics.
Research supported in part by the U.S. Department of Energy
under cooperative research agreement \# DE-FC02-94ER40818.

\appendix
\section{How to lasso a black hole}
\label{clasgen}
For a string worldsheet action that reads
\begin{eqnarray}
S &=& \frac{1}{2 \pi \alpha'} \int d^2 \sigma
(G_{\mu \nu} (\dot{X^{\mu}} \dot{X^{\nu}}
- {X^{\mu}}' {X^{\nu}}')
+B_{\mu \nu} ({X^{\mu}}' \dot{X^{\nu}}
- \dot{X^{\mu}} {X^{\nu}}')) \nonumber 
\end{eqnarray}
after gauge fixing, the string equations of motion are:
\begin{eqnarray}
\dot{\dot{X^{\mu}}}- {X^{\mu}}''+ {\Gamma^{\mu}}_{\rho \nu}
(\dot{X^{\rho}} \dot{X^{\nu}}
- {X^{\rho}}' {X^{\nu}}') + {H^{\mu}}_{\rho \nu}
 ({X^{\nu}}' \dot{X^{\rho}}
- \dot{X^{\nu}} {X^{\rho}}')
\label{eom}
\end{eqnarray}
where
\begin{eqnarray}
{\Gamma^{\mu}}_{\rho \nu} &=& \frac{1}{2} G^{\mu \alpha}
(G_{\alpha \nu,\rho} + G_{\alpha \rho,\nu}-G_{\rho \nu, \alpha})
\\
{H^{\mu}}_{\rho \nu} &=& \frac{1}{2} G^{\mu \alpha}
(B_{\rho \nu,\alpha}-B_{\alpha \nu, \rho} + B_{\alpha \rho,\nu}).
\end{eqnarray}
These equations should be supplemented with the constraint equations:
\begin{eqnarray}
G_{\mu \nu}  {X^{\mu}}' \dot{X^{\nu}} &=& 0 \nonumber \\
G_{\mu \nu} (\dot{X^{\mu}} \dot{X^{\nu}}
+ {X^{\mu}}' {X^{\nu}}') &=& 0.
\end{eqnarray}
Note in particular that a $\sigma$-independent solution for classical
strings is a geodesic in spacetime. (We made us of that fact
in section \ref{geodesics}.)
\subsection*{Winding strings}
As remarked in section \ref{gensol},
 in the BTZ background, we can obtain winding
strings by starting from geodesics, and adding a winding in $\phi$,
which is accompanied by a stretch of the string in the time direction
$t$. The new classical solution is given by:
\begin{eqnarray}
r &=& r_0(\tau) \nonumber \\
\phi &=& \phi_0(\tau) + n \sigma \nonumber \\
t &=& t_0(\tau) + n \tau,
\end{eqnarray}
where the original solution (labelled by $'0'$) represents a geodesic.
It is then easy to see that the new terms in the equations of motion
(\ref{eom})
are either linear or quadratic in $n$. The term linear in $n$ is
proportional to $({\Gamma^{\mu}}_{\nu t} + {H^{\mu}}_{\nu \phi})
\dot{X^{\nu}}$ (where $\mu, \nu$ are generic). 
This can be proven to vanish for the BTZ background
using the metric, the $H$-field ($H_{r \phi t} = r$),
and the Christoffel connection:
\begin{eqnarray}
{\Gamma^r}_{tt} &=& r f^2 \nonumber \\
{\Gamma^t}_{tr} &=& r f^{-2} \nonumber \\
{\Gamma^{\phi}}_{tr} &=& \frac{J}{2r} f^{-2} \nonumber \\
{\Gamma^r}_{rr} &=& \frac{J^2-4r^4}{4 r^3} f^{-2} \nonumber \\
{\Gamma^{t}}_{\phi r} &=& -\frac{J}{2r} f^{-2} \nonumber \\
{\Gamma^{\phi}}_{r \phi} &=& \frac{r^2-M}{ r} f^{-2} \nonumber \\
 {\Gamma^r}_{\phi \phi} &=& -r f^2, 
\end{eqnarray} 
where $f^2=r^2-M+\frac{J^2}{4 r^2}$.
Similarly, 
the term quadratic in $n$, proportional to 
${\Gamma^{\mu}}_{t t}-{\Gamma^{\mu}}_{\phi \phi} - 2 {H^{\mu}}_{\phi t}$
vanishes in the WZW background.
The constraint equations can be satisfied by making use of the internal
degrees of freedom (i.e. non-trivial dynamics in the other seven dimensions
of spacetime). The analysis in this appendix is similar in spirit
to the approach
of e.g. \cite{Larsen:1995af}\cite{deVega:1998ny}\cite{Larsen:2000yw}.

\section{$Sl(2,R)$ representation theory}
We are interested in representations of the Lie algebra of
generators of the group $SU(1,1)$, isomorphic to $Sl(2,R)$. We
do not restrict to proper $Sl(2,R)$ representations, but
want to include representations of the covering group
$\widetilde{SU(1,1)}$. Then the unitary
representations with Casimir $c_2=-\tau(\tau+1)$ are given by:
\begin{itemize}
\item the trivial representation $T_0$. 
\item the principal continuous representations $T_{(i \rho-\frac{1}{2},
\epsilon)}$, $\rho \in R$ and $\epsilon \in {\{}0, \frac{1}{2} {\}}$.
\item the principle discrete representations $T_l^+$ with lowest
weight $-l>0$ and $c_2=-l(l+1)$.
\item the principle discrete representations $T_l^-$ with 
highest weight $l<0$ and $c_2=-l(l+1)$.
\item the complementary representations $T_{(\tau,0)}$,
$-1 < \tau < 0$,
with lowest positive $J_0^3$
eigenvalue $ 0 \le \alpha <1$ for which we have the inequality
$ |\tau+\frac{1}{2}|<|\alpha-\frac{1}{2}|$. 
\end{itemize}
We refer to \cite{VK} and \cite{Dixon:1989cg} for more details.


\end{document}